\documentclass[twocolumn,a4paper,12pt]{article} 
\usepackage{epsfig}
\newcommand{\beq}{\begin{eqnarray}}
\newcommand{\eeq}{\end{eqnarray}}

\begin{document}
\pagestyle{plain}
%
%
\title{
{\Large \bf 
FEATHER: \\ a fast intra-pulse feedback system for the JLC.
}
}
 
\author{
{\bf 
Nicolas Delerue
} \\ 
JLC Group, KEK\\
Talk given at the 8th Accelerator and Particle Physics Institute
}

\date{
APPI (Japan), February 2003
}

\twocolumn[
\vspace*{-2cm}
\maketitle

\vspace*{-1cm}
%
\begin{center}
{\bf 
Ground motion at the Future JLC detector may affect beam alignment and
cause huge luminosity loss. The FEATHER (FEedback AT High Energy Requirements) project addresses this problem by designing a fast intra-pulse feedback
system that will correct the observed
beam offset.
} \\ 
\end{center}
]

%
\section{Need for a fast feedback system}
%

\begin{figure}[hbt]
\vspace{-.5cm}
\begin{tabular}{p{4cm}p{4cm}}
\centering\epsfig{file=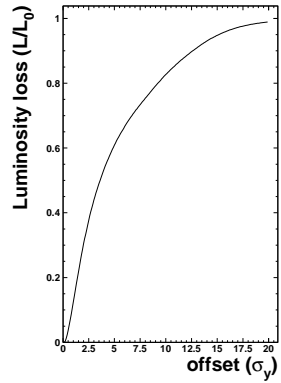,width=4.cm}&
\vspace*{-5.5cm}
\begin{tabular}{p{4cm}}
\caption{Fraction of the total luminosity lost  as a  function of the vertical offset of the beams at the interaction point. The horizontal unit, $\sigma_y$, is the vertical size of the beam (a few nanometers).}
\label{fig:lumiloss} 
\end{tabular}
\end{tabular}
\vspace{-.5cm}
\end{figure}

Ground motion arises from both natural and human activities. At frequencies higher than a few tens of hertz most of the motion comes from ``cultural noise'' (ie. human activities). At these frequencies the amplitude of the ground motion, a few nanometers, is comparable to the vertical beam size proposed for the JLC. This means that beams may be misaligned, leading to loss of luminosity (see figure~\ref{fig:lumiloss}) and poorer performances for the collider. The effects of beam misalignment have been studied using CAIN~\cite{Chen:1995jt}.

%
%

\begin{figure}[htb]
\epsfig{file=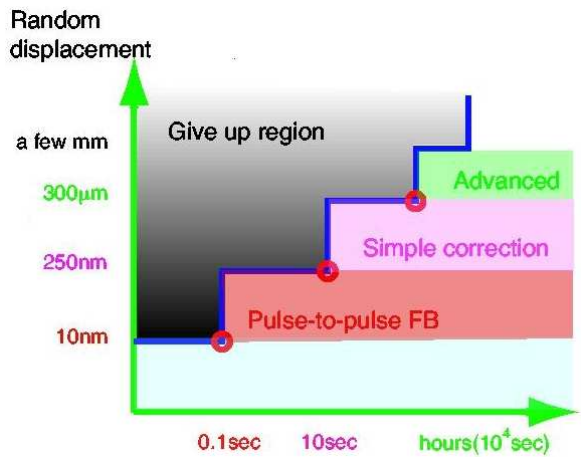,width=6.cm,angle=270}
\caption{Acceptable noise amplitude at various frequencies and methods that will be used to correct the resulting misalignment.}
\label{fig:gm} 
\end{figure}

This problem will partly be addressed during the site selection (see~\cite{uchida}) but it is unavoidable that some noise will remain (at least the noise arising from the accelerator operation). The figure~\ref{fig:gm} shows the acceptable noise amplitude for various frequency range. 

%
%

Ground motion will also be reduced by various mechanical device, but in the high frequency region, an active device is needed to completely correct the beam misalignment. This is the purpose of the fast feedback system proposed by the FEATHER (FEedback AT High Energy Requirement) collaboration~\cite{feather}.

%
\section{Different models of fast feedback system}
%

\begin{figure}[htbp]
\begin{tabular}{p{4cm}p{3cm}}
\centering{\epsfig{file=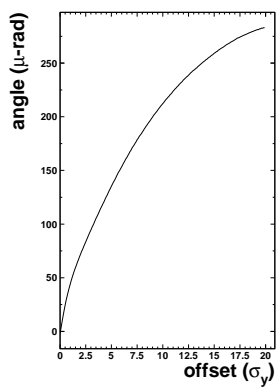,width=4.cm}} &
\vspace*{-4.5cm}
\begin{tabular}{p{3cm}}
\caption{Relation between the incoming offset and the deflected angle after the IP.}
\label{fig:angle} 
\end{tabular}
\end{tabular}
\end{figure}

As the beam travels at the speed of the light, it is not possible to use the position of a bunch to correct the same beam. Furthermore, the beam size is of the order of a few nanometers, and most displacements are well beyond the reach of current Beam Position Monitors (BPM). But after the interaction point (IP) both beams are deflected with a deflection angle defined by their misalignment (see figure~\ref{fig:angle}). Thus it is easier to measure the misalignment on the outgoing beam after the IP to correct the incoming beam. This can be done either on only one arm or on the two arms of the collider. The two arms solution reduces the correction required but is more complicated as good and fast communication between the two systems is required to avoid ``opposite'' corrections.

%
%
%

\subsection{Simple model of feedback}

\begin{figure}[hbtp]
\centering\epsfig{file=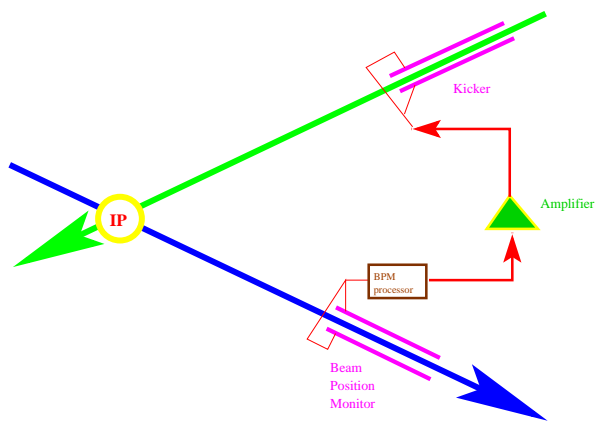,width=8.cm}
\caption{Layout of a simple model of intra-pulse feedback system.}
\label{fig:simpleLayout} 
\end{figure}

A simple system of feedback just needs to read the outgoing beam position from the BPM, compute the correction needed and then apply it to the incoming beam with a kicker. The layout of such system can be seen on figure~\ref{fig:simpleLayout}.

%
%

\begin{figure}[h]
\centering\epsfig{file=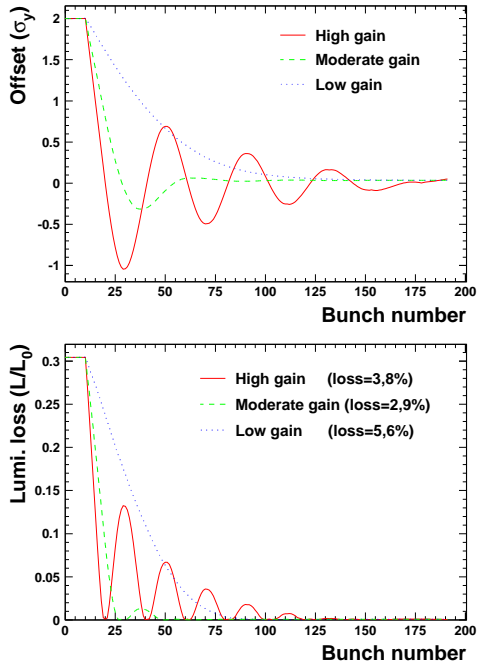,width=7.cm}
\caption{Offset (top) and luminosity loss (bottom) as a function of the bunch number in a train corrected using a simple model of feedback with a slow rise time kicker and an initial offset of 2 $\sigma_y$.}
\label{fig:simpleModelSlow} 
\end{figure}

Earlier studies of simple model of feedback have been done in 1999 \cite{Schulte:1999jn,Schulte:2000ax}. These studies used a kicker with a slow rise time (ie. high capacity), different from what is usually used at KEK.  The beam offset as a function of the bunch number as calculated in these studies can be seen on figure~\ref{fig:simpleModelSlow}.

%
%

As the kicker will have a slow rise time there will be an initial rising period during which the kicker can not fully correct the beam offset. With a high gain (red line), when full correction is achieved, the beam reaching the BPM is still not fully corrected and thus the system will ``overcorrect'' the beam position, leading to oscillations as seen on figure~\ref{fig:simpleModelSlow}. This can be avoided by lowering the gain of the correcting device, overcorrection will then be avoided but converging time will be much slower (blue dotted line). An average solution minimizing the luminosity loss is to use a moderate gain that will lead to a small overcorrection (green dashed line).

\begin{figure}[hbt]
\centering\epsfig{file=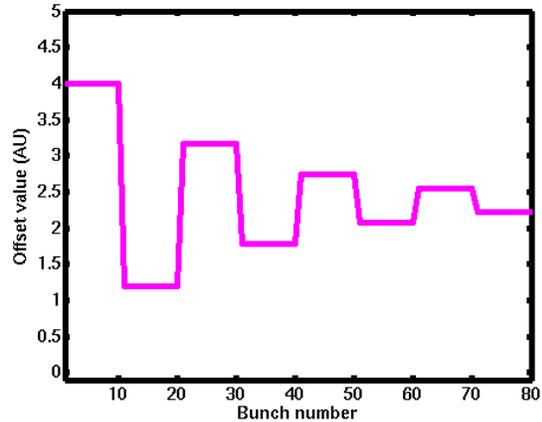,width=6.cm,angle=270}
\caption{Beam offset as a function of the bunch number in a train corrected with a simple model of feedback using a fast rise time kicker. Vertical units are arbitrary. After a few oscillations the system converges to an equilibrium state different from the full correction state.}
\label{fig:simpleModelFast} 
\end{figure}

This is modified if a fast rise time kicker (as those available at KEK) is used. The figure~\ref{fig:simpleModelFast} shows the beam offset as a function of the bunch number in that case.

%
%

Now, we can see that once the first bunches reach the BPM, correction is applied to the beam to reduce its offset. But when the corrected bunches (with a small offset) reach the BPM, the correction applied becomes small (or even 0 if the bunches are fully corrected). The system will then oscillate between high and low correction state. If the gain applied is lower than 1, the system will eventually converge to an equilibrium state different from the full correction state.

Animations showing the behavior of this model with these 2 kinds of kicker can be seen on the FEATHER website~\cite{feather}.

\subsection{Delayed model of feedback}

To avoid the oscillations featured by the ``simple model'', a ``memory'' can be added so that when corrected bunches reach the BPM, the previously applied correction is remembered and can be again applied.

\begin{figure}[htbp]
\centering\epsfig{file=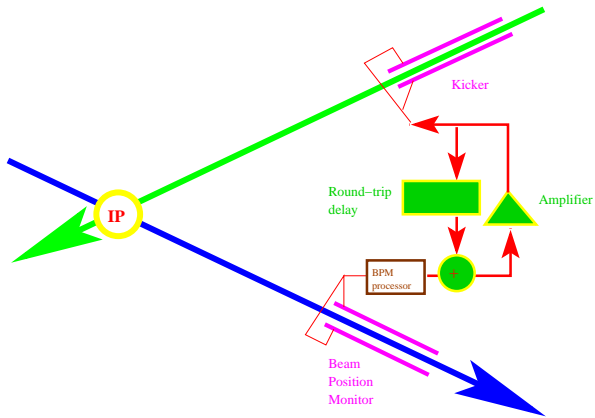,width=8.cm}
\caption{Layout of a model of intra-pulse feedback system using a delay line.}
\label{fig:delayedLayout} 
\end{figure}

\begin{figure}[hbt]
\centering\epsfig{file=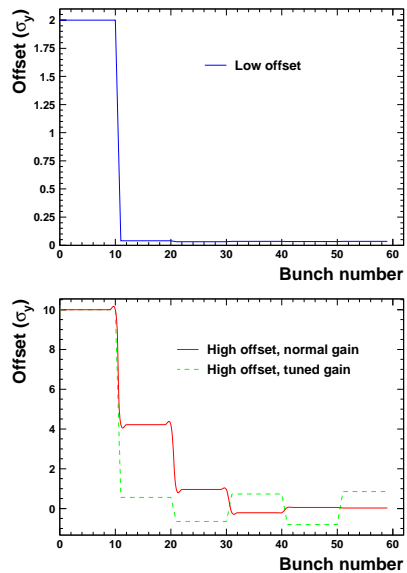,width=6.cm}
\caption{Offset as a function of the bunch number with a feedback system including a delay line. The upper plot is for a moderate initial offset (2 $\sigma_y$) for which the gain was tuned. The lower plot is for two different gains with high initial offset (10 $\sigma_y$), one (in red, plain curve) tuned for low offset (same gain as above) whereas the second one (in green, dotted line) is tuned for higher gains and shows instabilities at low offset.}
\label{fig:delayedModel} 
\end{figure}

The memory needs to remember the correction applied around 10 bunches ago, which means keeping the information just during a few tens of nanoseconds. This can be done simply by adding a long wire in the circuit as a ``delay'' loop (see figure~\ref{fig:delayedLayout}). The beam position as a function of the bunch number can be seen on figure~\ref{fig:delayedModel}.

%
%
%

%
%

In this delayed feedback, once the corrected bunches reach the BPM, the delay loop brings the information on the previously applied correction, avoiding the oscillations previously observed. But the choice of the correction to be applied as function of a given offset will trigger another problem as the relation between the position measured and the beam offset is non linear. If the correction is tuned for low offsets the high offsets will be undercorrected (red line on the bottom plot of figure~\ref{fig:delayedModel}). On the other hand, tuning the system for high offsets would lead to overcorrection and oscillations at low offset  (dashed green line on the bottom plot of figure~\ref{fig:delayedModel}).

\subsection{Improved model of feedback}

\begin{figure}[!h]
\begin{tabular}{p{4.3cm}p{2.5cm}}
\hspace*{-1cm}\epsfig{file=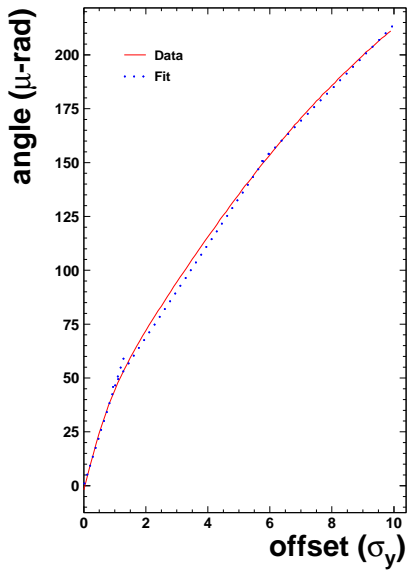,width=6.cm} &
\vspace*{-7.5cm}
\begin{tabular}{p{2.5cm}}
\caption{Fit (in blue, dotted line) of the simulated values (in red, plain line) of the deflected angle as function of the vertical beam offset at the IP.}
\label{fig:angle_fit} 
\end{tabular}
\end{tabular}
\end{figure}

\begin{figure}[hbt]
\centering\epsfig{file=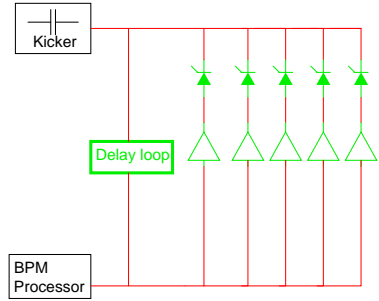,width=5.cm}
\caption{An array of non-linear components can be used to generate a non linear response.}
\label{fig:circuit_nonlineararray} 
\end{figure}

The delayed model of feedback system can be further improved to suppress the problem arising from the non linearities by fitting lines on the figure~\ref{fig:angle} showing the relation between offset and deflection angle.  This fit is shown on figure~\ref{fig:angle_fit}. It can be realized with an array of non linear components starting to operate at a different threshold as shown on figure~\ref{fig:circuit_nonlineararray}.

%
%
%

%
%
%

%
\section{Bench and beam tests}
%

\subsection{Feasibility of the 3 models}

The plots shown in the previous section have been made using numerical simulations in Perl (and later cross-checked with Matlab). It has not yet been possible to test these models in real beam conditions, however, the technical feasibility of these circuits has been checked on a test bench using a pulse generator.

\subsubsection{Simple feedback}

\begin{figure}[hbtp]
\begin{tabular}{p{3cm}p{3cm}}
\centering\epsfig{file=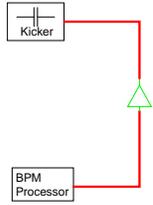,width=2.cm} &
\vspace*{-3cm}
\begin{tabular}{p{3cm}}
\caption{Electronic design of the simple model using a tunable amplifier.}
\label{fig:circuit_simple} 
\end{tabular}
\end{tabular}
\end{figure}

\begin{figure}[hbtp]
\centering\epsfig{file=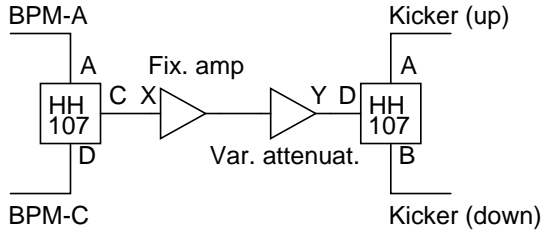,width=3.cm,angle=270}
\caption{Electronic design of the simple model where the tunable amplifier is replaced by a fixed one and a tunable attenuator. BPM and Kicker processing components are also shown.}
\label{fig:circuit_simpleRF} 
\end{figure}

As the simple model of feedback just computes a correction from a given position, its electronic layout is fairly simple: as shown on figure~\ref{fig:circuit_simple} it should consist simply of an amplifier whose gain is adjusted to the required correction. As tunable amplifier are not easily available on the market, it is easier to amplify more than needed and then tune the gain with well chosen attenuators as shown on figure~\ref{fig:circuit_simpleRF}. On that figure, the device used to merge the signal from the BPM antenna and to split it to both kicker strips is also shown.

%
%
%

%
%

\begin{figure}[!h]
\vspace*{-1cm}
\centering\epsfig{file=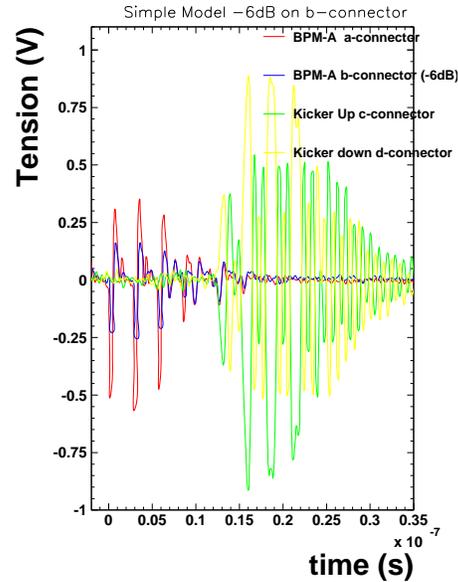,width=6.cm}
\caption{Test of the electronic circuit proposed for the simple model. In red and blue are the signal sent on the 2 BPM antennas of the circuit and in yellow and green are the responses sent to the strips of the kicker.}
\label{fig:simple_meas} 
\end{figure}

This design has been tested and its performances can be seen on figure~\ref{fig:simple_meas}. The response time measured is of the order of 15~ns.

%
%

\subsubsection{Delayed feedback}

\begin{figure}[hbtp]
\centering\epsfig{file=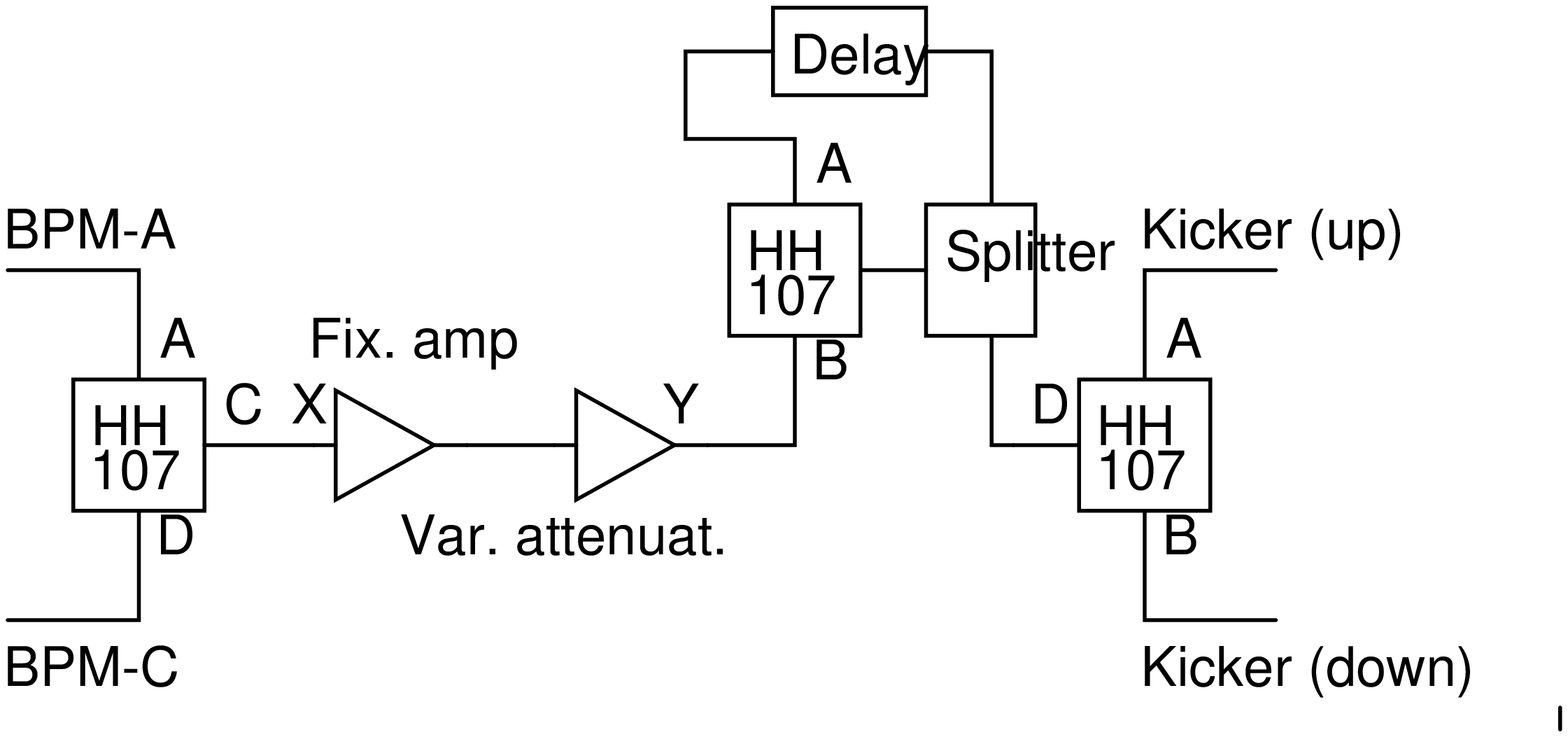,width=3.cm,angle=270}
\caption{Electronic design of the delayed model.}
\label{fig:circuit_delayedRF} 
\end{figure}

The delay loop to be added to the ``delayed feedback'' circuit could be made of a simple cable, but as there are losses, an amplifier must be included in the design to ensure that the delay loop has a gain of 1. The noise figure of this amplifier must be low to avoid accumulating and amplifying noise in the loop. The circuit layout is shown on figure~\ref{fig:circuit_delayedRF}.

%
%

\begin{figure}[!tb]
\vspace*{-1cm}
\centering\epsfig{file=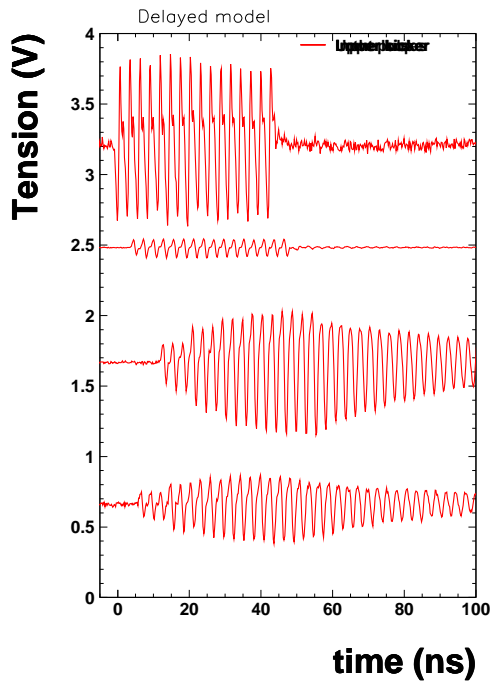,width=7.cm}
\caption{Test of the electronic circuit proposed for the delayed model. Vertical units are arbitrary. The upper line corresponds to an incoming train of bunches. The second line corresponds to the difference between the 2 BPM antennas, the third to the signal in the delay loop and the fourth to the signal sent to the kicker. The length of the delay loop is equivalent to the interval between 3 pulses, thus the signal in the delay loop increases by steps of 3 pulses. The decay in the delay after the end of incoming train comes from the fact that the gain in the loop is lower than 1.}
\label{fig:delayed_meas} 
\end{figure}

The figure~\ref{fig:delayed_meas} shows the input and output that were measured with this layout, using a delay loop with a length equivalent to 3 pulses. It can clearly be seen that the information is ``accumulated'' in the delay loop while bunches arrives with offset (which means that the correction is not yet enough) and then that once full correction is reached (no more signal on the input) almost the same level of correction is kept (the decay comes from the fact that the gain of the delay loop was not matched to one).

%
%

\subsubsection{Improved feedback: Non linearity at RF frequencies.}

A key issue to check the feasibility of the improved model of feedback is the possibility of generating a non linear response out of the components available at the working frequency. 

\begin{figure}[hbtp]
\epsfig{file=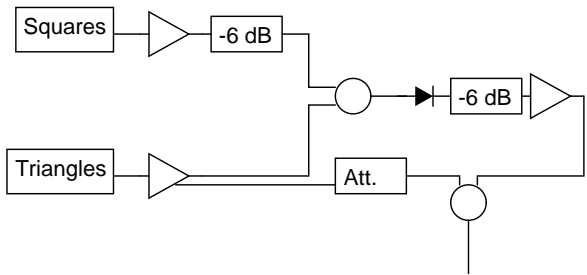,width=3.7cm,angle=270}
\caption{To have a ``threshold'' device at high frequency, a diode with shifted ground value must be used. The ground value is shifted by the addition, and later the substraction of a DC component.} 
\label{fig:circuit_nonlinear} 
\end{figure}

\begin{figure}[hbtp]
\epsfig{file=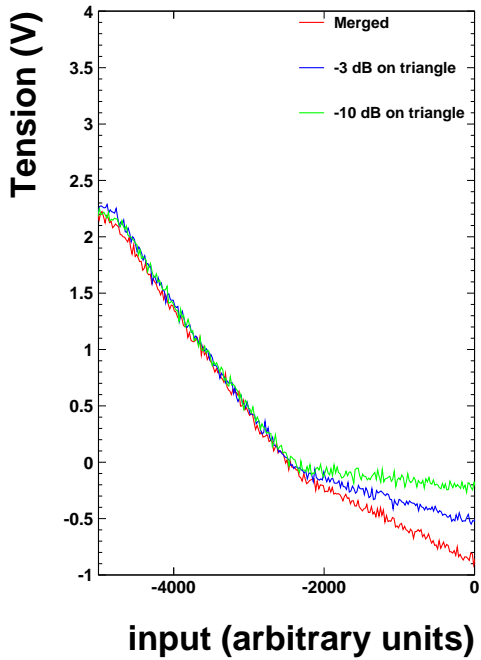,width=7.cm}
\caption{By using diode with a shifted ground value, it is possible to produce non linearities with linear input (here the input was a triangle signal).} 
\label{fig:nonlinear} 
\end{figure}

This non-linearity can be simulated by using a diode. The blocking tension of this diode can by shifted by adding a DC (or low frequency) component to the incoming signal and subtracting this DC components after the diode, as shown on figure~\ref{fig:circuit_nonlinear}. Using this design, a kink of various angle has been observed when a normal triangle signal was sent through the circuit (see figure~\ref{fig:nonlinear} ).

%
%
%

%
%

\subsection{Kicker requirements and signal filtering}

With the kickers currently available at KEK the tension required to kick the beam is of the order of the kilovolt. The only available amplifiers able to deliver this power does not work in the hundreds of megahertz range and thus the signal must be brought to a lower frequency.

\begin{figure}[htbp]
\epsfig{file=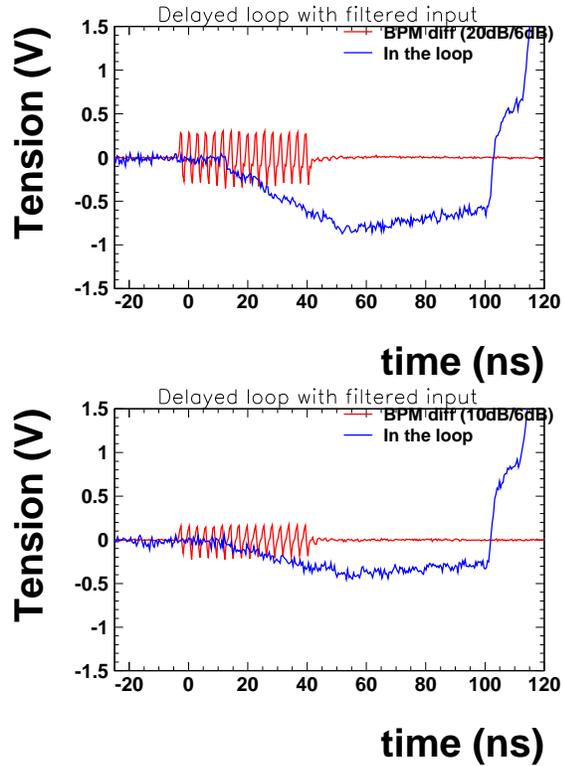,width=8.cm}
\caption{Delay loop with a lower frequency input. The 2 figures are for different beam position. The red curve shows the difference between the 2 BPM antenna and the blue curves shows the response of the feedback model: As long as the difference between the 2 BPM antennas is non zero the systems adjusts the BPM position. Once zero is reached, the achieved correction is kept by the delay loop (with a little decay) until the reset.} 
\label{fig:filteredDelayLoop} 
\end{figure}

This can be done by using a specific filter that reduces the frequency of the signal. It has been checked that the delayed model design remains valid with such filter (see figure~\ref{fig:filteredDelayLoop}).

%
%

\subsection{Online measurements}

\begin{figure}[htb]
\epsfig{file=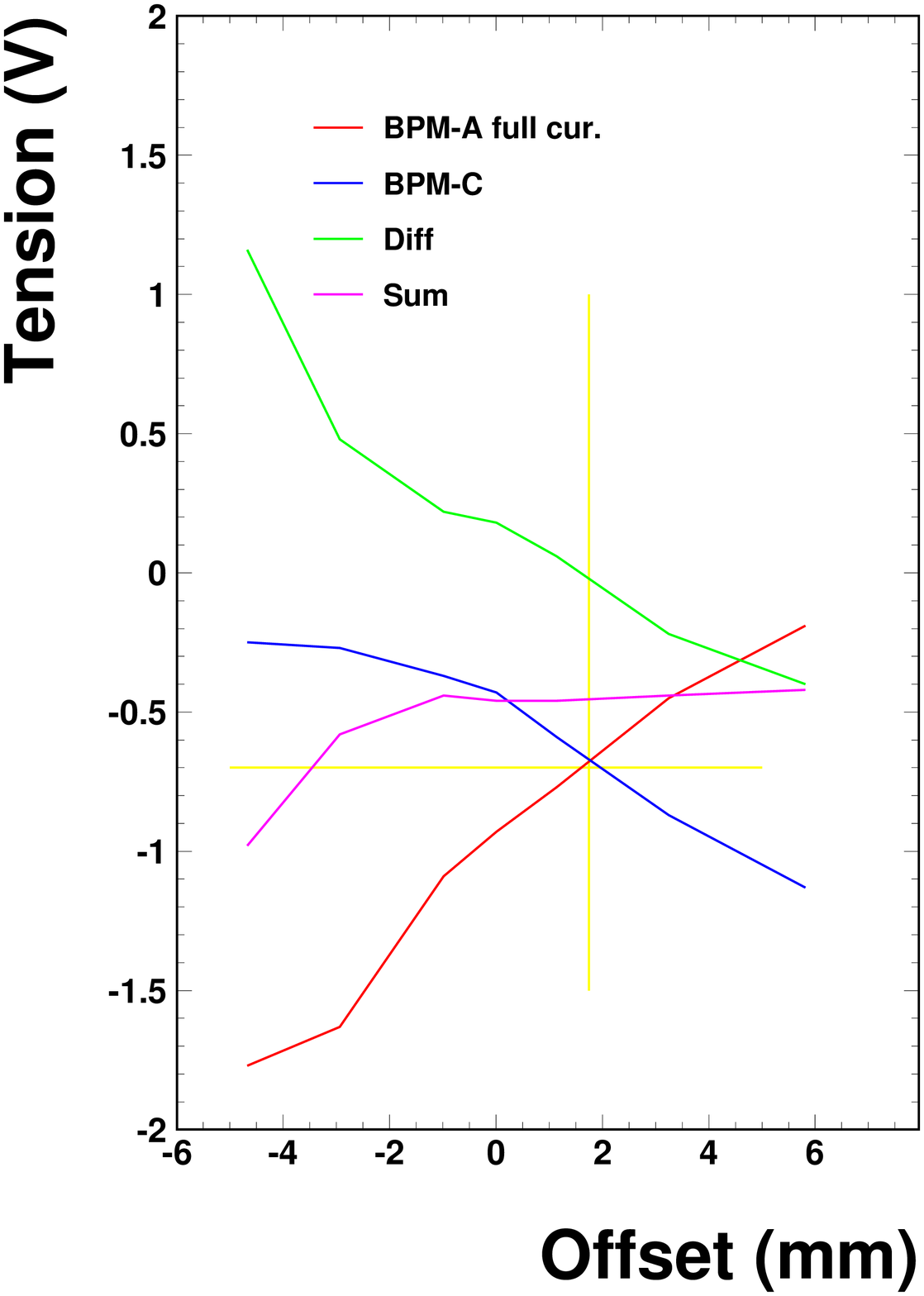,width=7.5cm}
\caption{Tension measured by the BPM as a function of the beam offset. The yellow line shows the axis with coordinates relative to the point where the upper and lower signal are equal (``zero'' of the BPM).}
\label{fig:offset_meas} 
\end{figure}

The response of a BPM as a function of the Beam Position has been measured at the ATF (here a ``button'' type BPM was used) as shown on figure~\ref{fig:offset_meas}.

%
%

\begin{figure}[!h]
\vspace*{-1.3cm}
\epsfig{file=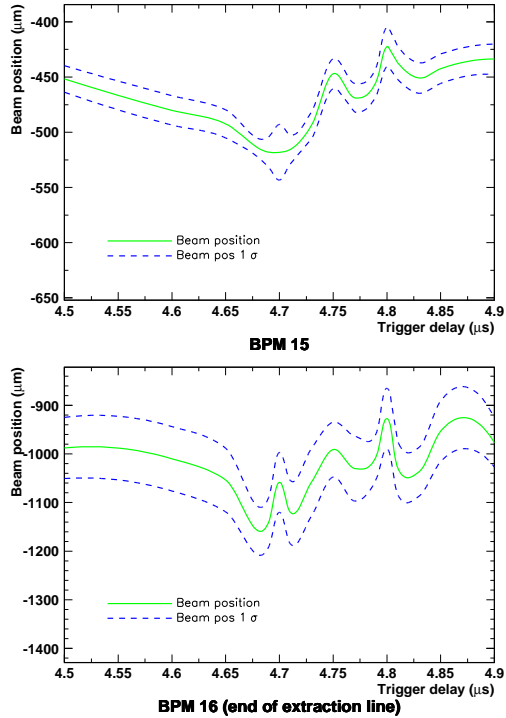,width=7.5cm}
\vspace*{-.3cm}
\caption{Position of the beam measured at 2 different BPM as a function of the delay between the ATF clock signal and the pulse sent to the kicker. A peak centered around 4.7~$\mu$s can be seen, it corresponds to the kick given to the beam. The measured value is given in green and the $\pm 1\sigma$ error limits are given in blue.}
\label{fig:kick_meas} 
\end{figure}

Other beam tests have been done to measure how the trajectory is modified by a given kick. Initial results were not conclusive due to insufficient power but the effect of the kick was later observed with a more powerful pulse generator (see figure~\ref{fig:kick_meas}). To be able to check the full feedback system a new kicker requiring less power needs to be designed.

%
%
%

%
\section{In the detector}
%

\begin{figure}[htbp]
\centering\epsfig{file=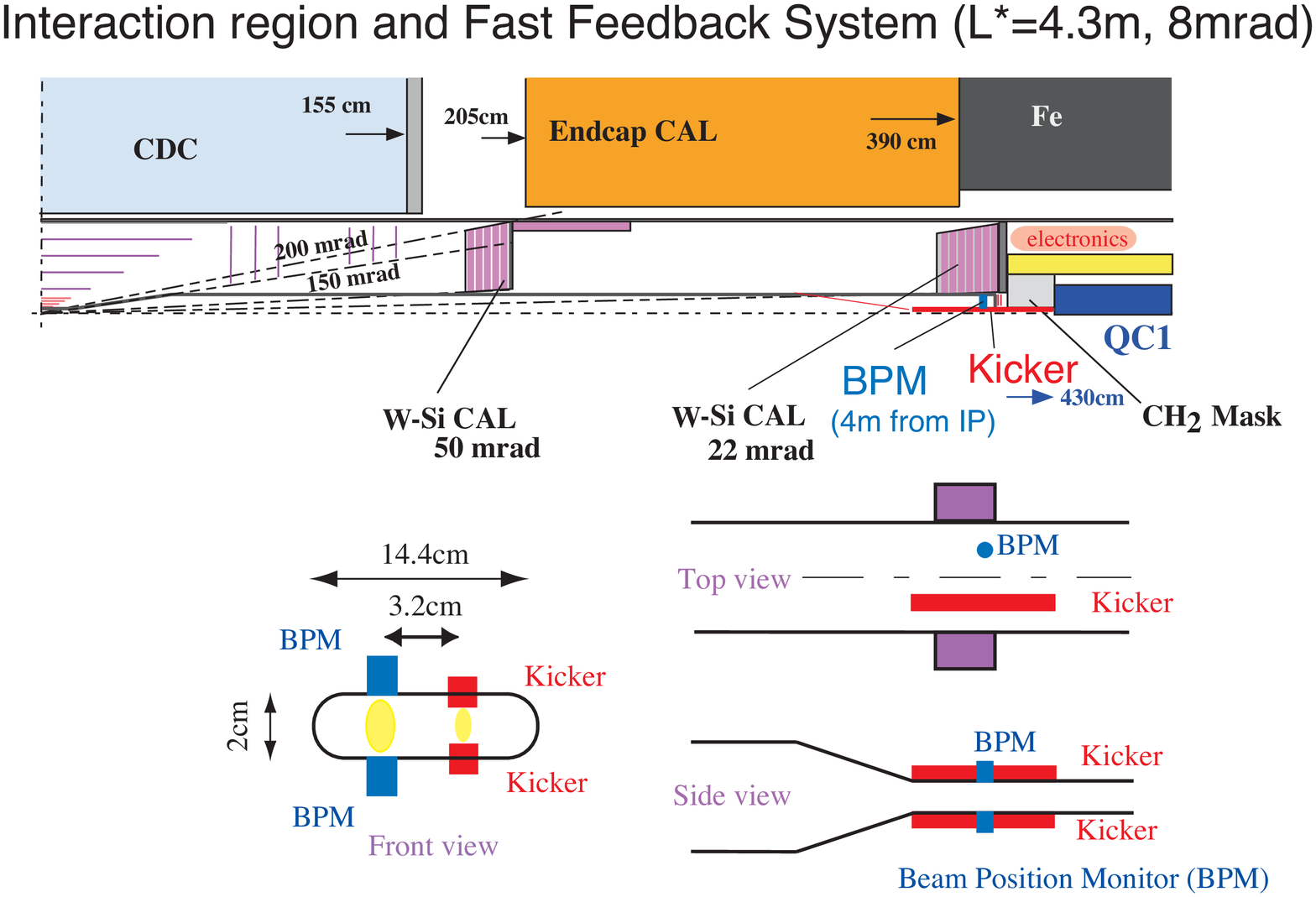,width=8.cm}
\caption{The FEATHER system is located 4m away from the IP in the detector. The top view shows how the system will be integrated in the detector, the 3 bottom views shown the configuration of the kicker and the BPM on the beam pipe. At the chosen location, both beams are still traveling in the same pipe.}
\label{fig:location} 
\end{figure}

The location of the feedback system will greatly influence its efficiency. The closest from the IP, the better performances can be reached. On an other hand, a system close from the IP will  suffer from high radiation rate and absorb valuable particles, affecting the detector's resolution. The best location seems to be 4 meters away from the IP as shown on figure~\ref{fig:location}.

%
%
%

%
\section{Conclusion}
%

The feasibility of a fast feedback system has been checked by the FEATHER collaboration. A new kicker is currently been designed. Once this new kicker will be ready new beam tests will be performed to confirm the results obtained on a test bench.

\bibliographystyle{myunsrt}
\bibliography{biblio}


ENTRY
  { address
    author
    booktitle
    chapter
    edition
    editor
    howpublished
    institution
    journal
    key
    month
    note
    number
    organization
    pages
    publisher
    school
    series
    title
    type
    volume
    year
  }
  {}
  { label }

INTEGERS { output.state before.all mid.sentence after.sentence after.block }

FUNCTION {init.state.consts}
{ #0 'before.all :=
  #1 'mid.sentence :=
  #2 'after.sentence :=
  #3 'after.block :=
}

STRINGS { s t }

FUNCTION {output.nonnull}
{ 's :=
  output.state mid.sentence =
    { ", " * write$ }
    { output.state after.block =
	{ add.period$ write$
	  newline$
	  "\newblock " write$
	}
	{ output.state before.all =
	    'write$
	    { add.period$ " " * write$ }
	  if$
	}
      if$
      mid.sentence 'output.state :=
    }
  if$
  s
}

FUNCTION {output}
{ duplicate$ empty$
    'pop$
    'output.nonnull
  if$
}

FUNCTION {output.check}
{ 't :=
  duplicate$ empty$
    { pop$ "empty " t * " in " * cite$ * warning$ }
    'output.nonnull
  if$
}

FUNCTION {output.bibitem}
{ newline$
  "\bibitem{" write$
  cite$ write$
  "}" write$
  newline$
  ""
  before.all 'output.state :=
}

FUNCTION {fin.entry}
{ add.period$
  write$
  newline$
}

FUNCTION {new.block}
{ output.state before.all =
    'skip$
    { after.block 'output.state := }
  if$
}

FUNCTION {new.sentence}
{ output.state after.block =
    'skip$
    { output.state before.all =
	'skip$
	{ after.sentence 'output.state := }
      if$
    }
  if$
}

FUNCTION {not}
{   { #0 }
    { #1 }
  if$
}

FUNCTION {and}
{   'skip$
    { pop$ #0 }
  if$
}

FUNCTION {or}
{   { pop$ #1 }
    'skip$
  if$
}

FUNCTION {new.block.checka}
{ empty$
    'skip$
    'new.block
  if$
}

FUNCTION {new.block.checkb}
{ empty$
  swap$ empty$
  and
    'skip$
    'new.block
  if$
}

FUNCTION {new.sentence.checka}
{ empty$
    'skip$
    'new.sentence
  if$
}

FUNCTION {new.sentence.checkb}
{ empty$
  swap$ empty$
  and
    'skip$
    'new.sentence
  if$
}

FUNCTION {field.or.null}
{ duplicate$ empty$
    { pop$ "" }
    'skip$
  if$
}

FUNCTION {emphasize}
{ duplicate$ empty$
    { pop$ "" }
    { "{\em " swap$ * "}" * }
  if$
}

INTEGERS { nameptr namesleft numnames }

FUNCTION {format.names}
{ 's :=
  #1 'nameptr :=
  s num.names$ 'numnames :=
  numnames 'namesleft :=
    { namesleft #0 > }
    { s nameptr "{ff~}{vv~}{ll}{, jj}" format.name$ 't :=
      nameptr #1 >
	{ namesleft #1 >
	    { ", " * t * }
	    { numnames #2 >
		{ "," * }
		'skip$
	      if$
	      t "others" =
		{ " et~al." * }
		{ " and " * t * }
	      if$
	    }
	  if$
	}
	't
      if$
      nameptr #1 + 'nameptr :=
      namesleft #1 - 'namesleft :=
    }
  while$
}

FUNCTION {format.authors}
{ author empty$
    { "" }
    { author format.names }
  if$
}

FUNCTION {format.editors}
{ editor empty$
    { "" }
    { editor format.names
      editor num.names$ #1 >
	{ ", editors" * }
	{ ", editor" * }
      if$
    }
  if$
}

FUNCTION {format.title}
{ title empty$
    { "" }
    { title "t" change.case$ }
  if$
}

FUNCTION {n.dashify}
{ 't :=
  ""
    { t empty$ not }
    { t #1 #1 substring$ "-" =
	{ t #1 #2 substring$ "--" = not
	    { "--" *
	      t #2 global.max$ substring$ 't :=
	    }
	    {   { t #1 #1 substring$ "-" = }
		{ "-" *
		  t #2 global.max$ substring$ 't :=
		}
	      while$
	    }
	  if$
	}
	{ t #1 #1 substring$ *
	  t #2 global.max$ substring$ 't :=
	}
      if$
    }
  while$
}

FUNCTION {format.date}
{ year empty$
    { month empty$
	{ "" }
	{ "there's a month but no year in " cite$ * warning$
	  month
	}
      if$
    }
    { month empty$
	'year
	{ month " " * year * }
      if$
    }
  if$
}

FUNCTION {format.btitle}
{ title emphasize
}

FUNCTION {tie.or.space.connect}
{ duplicate$ text.length$ #3 <
    { "~" }
    { " " }
  if$
  swap$ * *
}

FUNCTION {either.or.check}
{ empty$
    'pop$
    { "can't use both " swap$ * " fields in " * cite$ * warning$ }
  if$
}

FUNCTION {format.bvolume}
{ volume empty$
    { "" }
    { "volume" volume tie.or.space.connect
      series empty$
	'skip$
	{ " of " * series emphasize * }
      if$
      "volume and number" number either.or.check
    }
  if$
}

FUNCTION {format.number.series}
{ volume empty$
    { number empty$
	{ series field.or.null }
	{ output.state mid.sentence =
	    { "number" }
	    { "Number" }
	  if$
	  number tie.or.space.connect
	  series empty$
	    { "there's a number but no series in " cite$ * warning$ }
	    { " in " * series * }
	  if$
	}
      if$
    }
    { "" }
  if$
}

FUNCTION {format.edition}
{ edition empty$
    { "" }
    { output.state mid.sentence =
	{ edition "l" change.case$ " edition" * }
	{ edition "t" change.case$ " edition" * }
      if$
    }
  if$
}

INTEGERS { multiresult }

FUNCTION {multi.page.check}
{ 't :=
  #0 'multiresult :=
    { multiresult not
      t empty$ not
      and
    }
    { t #1 #1 substring$
      duplicate$ "-" =
      swap$ duplicate$ "," =
      swap$ "+" =
      or or
	{ #1 'multiresult := }
	{ t #2 global.max$ substring$ 't := }
      if$
    }
  while$
  multiresult
}

FUNCTION {format.pages}
{ pages empty$
    { "" }
    { pages multi.page.check
	{ "pages" pages n.dashify tie.or.space.connect }
	{ "page" pages tie.or.space.connect }
      if$
    }
  if$
}

FUNCTION {format.vol.num.pages}
{ volume field.or.null
  number empty$
    'skip$
    { "(" number * ")" * *
      volume empty$
	{ "there's a number but no volume in " cite$ * warning$ }
	'skip$
      if$
    }
  if$
  pages empty$
    'skip$
    { duplicate$ empty$
	{ pop$ format.pages }
	{ ":" * pages n.dashify * }
      if$
    }
  if$
}

FUNCTION {format.chapter.pages}
{ chapter empty$
    'format.pages
    { type empty$
	{ "chapter" }
	{ type "l" change.case$ }
      if$
      chapter tie.or.space.connect
      pages empty$
	'skip$
	{ ", " * format.pages * }
      if$
    }
  if$
}

FUNCTION {format.in.ed.booktitle}
{ booktitle empty$
    { "" }
    { editor empty$
	{ "In " booktitle emphasize * }
	{ "In " format.editors * ", " * booktitle emphasize * }
      if$
    }
  if$
}

FUNCTION {empty.misc.check}
{ author empty$ title empty$ howpublished empty$
  month empty$ year empty$ note empty$
  and and and and and
    { "all relevant fields are empty in " cite$ * warning$ }
    'skip$
  if$
}

FUNCTION {format.thesis.type}
{ type empty$
    'skip$
    { pop$
      type "t" change.case$
    }
  if$
}

FUNCTION {format.tr.number}
{ type empty$
    { "Technical Report" }
    'type
  if$
  number empty$
    { "t" change.case$ }
    { number tie.or.space.connect }
  if$
}

FUNCTION {format.article.crossref}
{ key empty$
    { journal empty$
	{ "need key or journal for " cite$ * " to crossref " * crossref *
	  warning$
	  ""
	}
	{ "In {\em " journal * "\/}" * }
      if$
    }
    { "In " key * }
  if$
  " \cite{" * crossref * "}" *
}

FUNCTION {format.crossref.editor}
{ editor #1 "{vv~}{ll}" format.name$
  editor num.names$ duplicate$
  #2 >
    { pop$ " et~al." * }
    { #2 <
	'skip$
	{ editor #2 "{ff }{vv }{ll}{ jj}" format.name$ "others" =
	    { " et~al." * }
	    { " and " * editor #2 "{vv~}{ll}" format.name$ * }
	  if$
	}
      if$
    }
  if$
}

FUNCTION {format.book.crossref}
{ volume empty$
    { "empty volume in " cite$ * "'s crossref of " * crossref * warning$
      "In "
    }
    { "Volume" volume tie.or.space.connect
      " of " *
    }
  if$
  editor empty$
  editor field.or.null author field.or.null =
  or
    { key empty$
	{ series empty$
	    { "need editor, key, or series for " cite$ * " to crossref " *
	      crossref * warning$
	      "" *
	    }
	    { "{\em " * series * "\/}" * }
	  if$
	}
	{ key * }
      if$
    }
    { format.crossref.editor * }
  if$
  " \cite{" * crossref * "}" *
}

FUNCTION {format.incoll.inproc.crossref}
{ editor empty$
  editor field.or.null author field.or.null =
  or
    { key empty$
	{ booktitle empty$
	    { "need editor, key, or booktitle for " cite$ * " to crossref " *
	      crossref * warning$
	      ""
	    }
	    { "In {\em " booktitle * "\/}" * }
	  if$
	}
	{ "In " key * }
      if$
    }
    { "In " format.crossref.editor * }
  if$
  " \cite{" * crossref * "}" *
}

FUNCTION {article}
{ output.bibitem
  format.authors "author" output.check
  new.block
  format.title "title" output.check
  new.block
  crossref missing$
    { journal emphasize "journal" output.check
      format.vol.num.pages output
      format.date "year" output.check
    }
    { format.article.crossref output.nonnull
      format.pages output
    }
  if$
  new.block
  note output
  fin.entry
}

FUNCTION {book}
{ output.bibitem
  author empty$
    { format.editors "author and editor" output.check }
    { format.authors output.nonnull
      crossref missing$
	{ "author and editor" editor either.or.check }
	'skip$
      if$
    }
  if$
  new.block
  format.btitle "title" output.check
  crossref missing$
    { format.bvolume output
      new.block
      format.number.series output
      new.sentence
      publisher "publisher" output.check
      address output
    }
    { new.block
      format.book.crossref output.nonnull
    }
  if$
  format.edition output
  format.date "year" output.check
  new.block
  note output
  fin.entry
}

FUNCTION {booklet}
{ output.bibitem
  format.authors output
  new.block
  format.title "title" output.check
  howpublished address new.block.checkb
  howpublished output
  address output
  format.date output
  new.block
  note output
  fin.entry
}

FUNCTION {inbook}
{ output.bibitem
  author empty$
    { format.editors "author and editor" output.check }
    { format.authors output.nonnull
      crossref missing$
	{ "author and editor" editor either.or.check }
	'skip$
      if$
    }
  if$
  new.block
  format.btitle "title" output.check
  crossref missing$
    { format.bvolume output
      format.chapter.pages "chapter and pages" output.check
      new.block
      format.number.series output
      new.sentence
      publisher "publisher" output.check
      address output
    }
    { format.chapter.pages "chapter and pages" output.check
      new.block
      format.book.crossref output.nonnull
    }
  if$
  format.edition output
  format.date "year" output.check
  new.block
  note output
  fin.entry
}

FUNCTION {incollection}
{ output.bibitem
  format.authors "author" output.check
  new.block
  format.title "title" output.check
  new.block
  crossref missing$
    { format.in.ed.booktitle "booktitle" output.check
      format.bvolume output
      format.number.series output
      format.chapter.pages output
      new.sentence
      publisher "publisher" output.check
      address output
      format.edition output
      format.date "year" output.check
    }
    { format.incoll.inproc.crossref output.nonnull
      format.chapter.pages output
    }
  if$
  new.block
  note output
  fin.entry
}

FUNCTION {inproceedings}
{ output.bibitem
  format.authors "author" output.check
  new.block
  format.title "title" output.check
  new.block
  crossref missing$
    { format.in.ed.booktitle "booktitle" output.check
      format.bvolume output
      format.number.series output
      format.pages output
      address empty$
	{ organization publisher new.sentence.checkb
	  organization output
	  publisher output
	  format.date "year" output.check
	}
	{ address output.nonnull
	  format.date "year" output.check
	  new.sentence
	  organization output
	  publisher output
	}
      if$
    }
    { format.incoll.inproc.crossref output.nonnull
      format.pages output
    }
  if$
  new.block
  note output
  fin.entry
}

FUNCTION {conference} { inproceedings }

FUNCTION {manual}
{ output.bibitem
  author empty$
    { organization empty$
	'skip$
	{ organization output.nonnull
	  address output
	}
      if$
    }
    { format.authors output.nonnull }
  if$
  new.block
  format.btitle "title" output.check
  author empty$
    { organization empty$
	{ address new.block.checka
	  address output
	}
	'skip$
      if$
    }
    { organization address new.block.checkb
      organization output
      address output
    }
  if$
  format.edition output
  format.date output
  new.block
  note output
  fin.entry
}

FUNCTION {mastersthesis}
{ output.bibitem
  format.authors "author" output.check
  new.block
  format.title "title" output.check
  new.block
  "Master's thesis" format.thesis.type output.nonnull
  school "school" output.check
  address output
  format.date "year" output.check
  new.block
  note output
  fin.entry
}

FUNCTION {misc}
{ output.bibitem
  format.authors output
  title howpublished new.block.checkb
  format.title output
  howpublished new.block.checka
  howpublished output
  format.date output
  new.block
  note output
  fin.entry
  empty.misc.check
}

FUNCTION {phdthesis}
{ output.bibitem
  format.authors "author" output.check
  new.block
  format.btitle "title" output.check
  new.block
  "PhD thesis" format.thesis.type output.nonnull
  school "school" output.check
  address output
  format.date "year" output.check
  new.block
  note output
  fin.entry
}

FUNCTION {proceedings}
{ output.bibitem
  editor empty$
    { organization output }
    { format.editors output.nonnull }
  if$
  new.block
  format.btitle "title" output.check
  format.bvolume output
  format.number.series output
  address empty$
    { editor empty$
	{ publisher new.sentence.checka }
	{ organization publisher new.sentence.checkb
	  organization output
	}
      if$
      publisher output
      format.date "year" output.check
    }
    { address output.nonnull
      format.date "year" output.check
      new.sentence
      editor empty$
	'skip$
	{ organization output }
      if$
      publisher output
    }
  if$
  new.block
  note output
  fin.entry
}

FUNCTION {techreport}
{ output.bibitem
  format.authors "author" output.check
  new.block
  format.title "title" output.check
  new.block
  format.tr.number output.nonnull
  institution "institution" output.check
  address output
  format.date "year" output.check
  new.block
  note output
  fin.entry
}

FUNCTION {unpublished}
{ output.bibitem
  format.authors "author" output.check
  new.block
  format.title "title" output.check
  new.block
  note "note" output.check
  format.date output
  fin.entry
}

FUNCTION {default.type} { misc }

MACRO {jan} {"January"}

MACRO {feb} {"February"}

MACRO {mar} {"March"}

MACRO {apr} {"April"}

MACRO {may} {"May"}

MACRO {jun} {"June"}

MACRO {jul} {"July"}

MACRO {aug} {"August"}

MACRO {sep} {"September"}

MACRO {oct} {"October"}

MACRO {nov} {"November"}

MACRO {dec} {"December"}

MACRO {acmcs} {"ACM Computing Surveys"}

MACRO {acta} {"Acta Informatica"}

MACRO {cacm} {"Communications of the ACM"}

MACRO {ibmjrd} {"IBM Journal of Research and Development"}

MACRO {ibmsj} {"IBM Systems Journal"}

MACRO {ieeese} {"IEEE Transactions on Software Engineering"}

MACRO {ieeetc} {"IEEE Transactions on Computers"}

MACRO {ieeetcad}
 {"IEEE Transactions on Computer-Aided Design of Integrated Circuits"}

MACRO {ipl} {"Information Processing Letters"}

MACRO {jacm} {"Journal of the ACM"}

MACRO {jcss} {"Journal of Computer and System Sciences"}

MACRO {scp} {"Science of Computer Programming"}

MACRO {sicomp} {"SIAM Journal on Computing"}

MACRO {tocs} {"ACM Transactions on Computer Systems"}

MACRO {tods} {"ACM Transactions on Database Systems"}

MACRO {tog} {"ACM Transactions on Graphics"}

MACRO {toms} {"ACM Transactions on Mathematical Software"}

MACRO {toois} {"ACM Transactions on Office Information Systems"}

MACRO {toplas} {"ACM Transactions on Programming Languages and Systems"}

MACRO {tcs} {"Theoretical Computer Science"}

READ

STRINGS { longest.label }

INTEGERS { number.label longest.label.width }

FUNCTION {initialize.longest.label}
{ "" 'longest.label :=
  #1 'number.label :=
  #0 'longest.label.width :=
}

FUNCTION {longest.label.pass}
{ number.label int.to.str$ 'label :=
  number.label #1 + 'number.label :=
  label width$ longest.label.width >
    { label 'longest.label :=
      label width$ 'longest.label.width :=
    }
    'skip$
  if$
}

EXECUTE {initialize.longest.label}

ITERATE {longest.label.pass}

FUNCTION {begin.bib}
{ preamble$ empty$
    'skip$
    { preamble$ write$ newline$ }
  if$
  "\begin{thebibliography}{"  longest.label  * "}" * write$ newline$
}

EXECUTE {begin.bib}

EXECUTE {init.state.consts}

ITERATE {call.type$}

FUNCTION {end.bib}
{ newline$
  "\end{thebibliography}" write$ newline$
}

EXECUTE {end.bib}

\end{document}